\documentclass[12pt, a4paper]{article}

\usepackage{amsmath,amsfonts,amssymb,cite}
\usepackage[cp1251,ctt]{inputenc}
\usepackage[english]{babel}
\usepackage[T2A]{fontenc}

\let\Cal\mathcal
\let\cal\mathcal

\begin{document}
\title {
$${}$$
{\bf Non-singular screw dislocations as the Coulomb gas with smoothed out coupling and the renormalization of the shear modulus}{\footnote{ Presented at the International
Workshop ``Classical and Quantum Integrable Systems (CQIS-2011)''
(January 24-27, 2011, Protvino, Russian Federation)}}}
\author{
$${}$$
{\bf\Large C. Malyshev}\\
$${}$$\\
{\it\small Steklov Institute of Mathematics}
{\it\small (St.-Petersburg Department)}\\
{\it\small Fontanka 27, St.-Petersburg, 191023, RUSSIA}}

\date{}

\maketitle

\def \bbe{\boldsymbol\be}
\def \bchi{\boldsymbol\chi}
\def \bdl{\boldsymbol\dl}
\def \bphi{\boldsymbol\phi}
\def \bsi{\boldsymbol\si}
\def \bta{\boldsymbol\eta}
\def \bxi{\boldsymbol\xi}
\def \bna{\boldsymbol\nabla}
\def \bvphi{\boldsymbol\varphi}
\def \bPsi{\boldsymbol\Psi}
\def \binc{\boldsymbol\Inc}
\def \bx{\bf{x}}
\def \bQ{{\bf Q}}
\def \bcP{\boldsymbol{\cal P}}

\def \al{\alpha}
\def \be{\beta}
\def \ga{\gamma}
\def \dl{\delta}
\def \ze{\zeta}
\def \nb{\nabla}
\def \th{\theta}
\def \la{\lambda}
\def \si{\sigma}
\def \om{\omega}
\def \z{\zeta}
\def \Ga{\Gamma}
\def \Dl{\Delta}
\def \La{\Lambda}
\def \Si{\Sigma}
\def \Ph{\Phi}
\def \Om{\Omega}
\def \ph{\varphi}
\def \vt{\vartheta}
\def \ep{\varepsilon}
\def \ka{\varkappa}

\def \cA{\cal A}
\def \cB{\cal B}
\def \cC{\cal C}
\def \cD{\cal D}
\def \cE{\cal E}
\def \cF{\cal F}
\def \cG{\cal G}
\def \cI{\cal I}
\def \cJ{\cal J}
\def \cK{\cal K}
\def \cN{\cal N}
\def \cP{\cal P}
\def \cR{\cal R}
\def \cS{\cal S}
\def \cT{\cal T}
\def \cY{\cal Y}
\def \cZ{\cal Z}
\def \cM{{\cal M}}
\def \cL{{\cal L}}
\def \CU{{\cal U}}
\def \CW{{\cal W}}

\def \BC{\mathbb{C}}
\def \BD{\mathbb{D}}
\def \BZ{\mathbb{Z}}
\def \BR{\mathbb{R}}
\def \BQ{\mathbb{Q}}
\def \at{{\rm arctan}\,}
\def \ch{{\rm ch}\,}
\def \sh{{\rm sh}\,}
\def \th{{\rm th}\,}
\def \bg{{\rm bg}\,}
\def \Tr{{\rm Tr}\,}
\def \tr{{\rm tr}\,}
\def \Det{{\rm Det}\,}
\def \Inc{{\rm Inc}\,}
\def \diag{{\rm diag}\,}
\def \e{{\rm e}\,}
\def \c{{\rm c}\,}
\def \m{{\rm m}\,}
\def \d{{\rm d}}
\def \o{{\rm o}\,}

\def \w{\widetilde}
\def \h{\widehat}
\def \nt{{\widetilde n}}
\def \t{\times}
\def \r{\rangle}
\def \l{\langle}
\def \lav{\l\!\l}
\def \rav{\r\!\r}
\def \ld{\ldots}
\def \IM{\Im}
\def \RE{\Re}
\def \1{^{-1}}
\def \cd{\partial}

\vskip1.0cm
%%%%%%%%%%%%%%%%%%%%%%%%%%%%%%%%%%%%%%%%%%%%%%%%%%%%%%%%%%%%%%%%%%%%%%%%%

\begin{abstract}
\noindent
A field theory is developed for a
thermodynamical description of an array of parallel
non-singular screw dislocations in an elastic cylinder. The partition function of the system
is considered in the functional integral form.
Self-energy of the dislocation cores is chosen in the form suggested by the gauge-translational model of non-singular screw dislocation. It is shown that the system of the
dislocations is equivalent to the
two-dimensional Coulomb gas. The coupling potential is prevented from a
short-distance divergency since the core energies are taken into
account. Two-point
correlation functions of the stress components are obtained. Renormalization of
the shear modulus caused by the presence of the dislocations is studied in the approximation of non-interacting dislocation dipoles.
It is demonstrated that the finite size of the dislocation cores results in a modification of the renormalization law.

\end{abstract}


\thispagestyle{empty}
\newpage

\section{Introduction}
\label{sec1}

Non-trivial topologically configurations in ordered states (e.g. vortices,
dislocations, and other defects)
attract appreciable attention in modern condensed matter physics.
For instance, dislocations as imperfections of the crystalline
ordering are of importance for structural, transport, and
electronic properties of real solids.
Two-dimensional ordered states are of special interest
due to the significance of the topological defects for the corresponding phase
transitions \cite{ber1, ber2, kos1, kos2, pop1, pop2}. The ideas of \cite{ber1, ber2, kos1, kos2, pop1, pop2}
have been further elaborated for the description of the
dislocation-mediated crystal melting in two dimensions \cite{holz,
nel1, nel12, nel13, nel2}. The textbooks \cite{kl11, kl12} summarize an original approach to the
ordered states and phase transitions dominated by the line-like
disturbances. The peculiarity of \cite{kl11, kl12} is due to a
systematical usage of (singular) gauge fields. Certain aspects of
the statistical physics of dislocations are considered in \cite{kl2, yam,
rab, zan1, igg}.

Dislocations have recently attracted considerable attention as far as the physics of the nanotubes is concerned
\cite{saito, toman, g4, dkl}. The point is that multilayer nanotubes
can contain within their walls screw dislocations lying along the axis of the
tube \cite{g4}. The external diameter of the nanotubes ranges from nanometers to tens of nanometers. The wall thickness varies from one to several tens of atomic layers. Therefore, effects influenced by the dislocation cores look attractive for study in the context of the nanotubes since the core sizes are comparable with those of atomic layers. Note that the electronic and mechanical properties of the graphene sheets in presence of
dislocations are also of interest \cite{graph0, graph, graph1}.

According to the elasticity theory, the stress tensor components of a single dislocation are singular on the
defect line. In reality, the stress components are smoothed out because of the formation of the core regions. Since the first attempts \cite{pei, nab},
various approaches to dislocations with non-trivial core are
known. For instance, the quasi-continuum approach \cite{brail1,
kun}, the gradient elasticity \cite{cem1, g1}, and the Lagrangian
translational gauging \cite{val, ed1, mal1, laz2, laz3} enable one to
obtain appropriate solutions for non-singular screw dislocations.

The elastic stresses of the screw dislocation are
obtained in \cite{val, ed1, mal1, laz2, laz3} in the form of the
superposition of the conventional far-reaching  (''background'')
contribution and a short-ranged (``gauge'') correction. The latter
modifies the background fields within a compact core.
Therefore, smoothing of the singular behaviour of the conventional screw dislocation occurs
thus leading to the so-called \textit{modified} screw dislocation.
Note that the second-order elasticity is also of importance for the
defects in crystals \cite{haif}. The approach of \cite{mal1} is suitable for study of the second order stresses in the case of a non-singular screw dislocation \cite{mal2}.

This paper is concerned with a thermodynamical description of a large number of
the modified screw dislocations \cite{mal1} lying within a long
enough cylinder of a circular cross-section. The functional integration
\cite{pop3, pop4, riv, yr, kl3, zinn} is used to represent the partition
function and to study certain averages. The core energies are accounted for in accordance with \cite{mal1}. It is
shown that the collection of parallel non-singular dislocations is
equivalent to the two-dimensional Coulomb-like system of charges
interacting {\it via} the potential which is logarithmic at large distances
but vanishes locally (smoothed out coupling). The two-dimensional Coulomb gas \cite{deut} belongs to the class of
systems covered by the theory \cite{ber1, ber2, kos1, kos2, pop1, pop2}, and it demonstrates a relationship with the two-dimensional spin models
\cite{jos, nien, pol}.

Dislocations available in a solid result in the renormalization
of the corresponding elastic moduli \cite{nel12,
rab}. Two-point correlation functions of the stress components are calculated in this paper for investigating of the renormalization of the
shear modulus. It is demonstrated that
non-triviality of the cores results in a correction to the law of
the renormalization of the shear modulus what could be appreciable for the
nanotubes. The present field-theoretical approach is influenced
technically by that of \cite{ab, ab1}, where certain correlation
functions have been calculated for the string models possessing
the world-sheet vortices.

The paper is organized as follows. section~\ref{sec1} is the
introductory section. The partition function of the elastic cylinder containing the array of the modified screw dislocations is considered in section~\ref{sec2}. Transformation of collection of the dislocations to a dual system of two-dimensional charges is
presented in section~\ref{sec2}. In section~\ref{sec3}, we calculate certain
thermodynamical averages, e.g., the average square of the
dislocation dipole momentum and the stress-stress correlation
function in the approximation of non-interacting dislocation dipoles. The
renormalization of the shear modulus in the presence of non-singular
dislocations is studied section~\ref{sec4}. Discussion in section~\ref{sec5}
closes the paper.

\section{The partition function}
\label{sec2}

Consider a circular cylinder containing straight non-singular screw
dislocations lying along the axis of the cylinder. The
cylinder's material is approximated by elastically-isotropic
continuum described by linear elasticity. Recall that the screw
dislocation is characterized by the parallelism between its Burgers
vector and tangent to the dislocation line \cite{teod, hirth}. The
modified screw dislocation \cite{mal1} is a point of departure of
the present investigation that treats the distribution of the
dislocations as a thermodynamical ensemble at non-zero
temperature. The functional integration approach is used below to investigate the partition function and certain correlation functions. In what
follows, the Cartesian axis $Ox_3$ is along the cylinder's axis.

Let us begin with the functional integral representation of the thermodynamical partition function ${\cZ}$ of the
elastic cylinder containing the non-singular dislocations:
\begin{eqnarray}
&&{\cZ}\,=\,\frac1N\int e^{-\be W}\,\cD ({ \si}^{\rm b}_{i j},
{{\si}}^{\rm c}_{i j}, u_i, e_{i j})\,, \label{cor:partf01}
\\
&&W\,\equiv\,E- i E_{{\rm ext}}\,,\qquad
E\,\equiv\,E_{{\rm el}}+E_{{\rm core}}\,, \label{cor:partf02}
\end{eqnarray}
where $\be$ is inverse of the absolute temperature $T$ (the Boltzmann constant is unity). The functional $W$ (\ref{cor:partf02}) consists of the following contributions (indices repeated imply summation):
\begin{equation}
\begin{array}{rcl}
&&E_{\rm el}\,=\,\displaystyle{\frac{1}{4\mu}\int
\bigl(({\si}^{\rm b}_{i j} + {\si}^{\rm c}_{i j})^2 -
\frac{\nu}{1+\nu} ({\si}^{\rm b}_{i i} +
{\si}^{\rm c}_{i i})^2\bigr)\,\d^3x}\,,\\
&&E_{\rm core}\,=\,\displaystyle{\int (\ell\, e_{i j}({\text{inc}}\,e)_{i j}\,-\,e_{i j}\, {\si}^{\rm c}_{i j})\,\d^3x }\,, \\
&&E_{\rm ext}\,=\,\displaystyle{\frac12\int {\si}^{\rm b}_{i j}
(\cd_i u_j + \cd_j u_i- 2{\cP}_{i j})\,\d^3x}\,,
\end{array}
 \label{cor:en1}
\end{equation}
where $({\text{inc}}\,e)_{i j}$ denotes the double-curl operator: $ - {\epsilon}_{i k l} {\epsilon}_{j m n} \cd_k\,\cd_m e_{l n}$ (${\epsilon}_{i k l}$ is
totally antisymmetric tensor).
The functional $E_{\rm el}$ (\ref{cor:en1}) is the elastic energy
of superposition of two stresses, ${\si}^{\rm b}_{i j}$ and
${\si}^{\rm c}_{i j}$, provided $\mu$ and $\nu$ are respectively
identified as the shear modulus and the Poisson ratio. The
notation ${\si}^{\rm b}_{i j}$ is reserved for the long-ranged
contribution, while ${\si}^{\rm c}_{i j}$ describes a
non-conventional stress which modifies the background one,
${\si}^{\rm b}_{i j}$, within the core of the modified dislocation
\cite{mal1}~\footnote { Two stress fields ${\si}^{\rm b}_{i j}$
and ${\si}^{\rm c}_{i j}$ correspond to splitting of the total
elastic stress into, so-called, ``background'' and ``gauge'' parts
\cite{ed1, mal1, laz2, laz3}.}.  The dislocation core energy is given by $E_{\rm core}$ (\ref{cor:en1}), where
$e_{i j} ({\text{inc}}\,e)_{i j}$  originates (by means of
linearizations) from the Hilbert--Einstein Lagrangian
proposed in \cite{mal1} in the framework of the translational gauge
approach to dislocations~\footnote {The gauge Lagrangians
quadratic in the dislocation densities are used in \cite{ed1,
laz2} what is equivalent, in the case of the screw dislocation, to
the approach of \cite{mal1}. The most general three-dimensional gauge Lagrangian has been discussed in \cite{vol}. Certain developments in the gauge gravity \cite{kl12, sard, hehl, kl4, katan} involve the gauge-translational features of the theory of dislocations.}.
Here $e_{i
j}$ is the total strain tensor, and
$\ell$ is a scale of the dislocation core energy. The term
$E_{\rm ext}$ (\ref{cor:en1}) is linear with respect to
(symmetrized) derivatives of the displacement vector $u_i$, as
well as with respect to a ``source'' $\cP_{i j}$ which is related
to the plastic strain $e^P_{i j}$ as follows: $\cP_{i j} = e^P_{i
j}+C_{i j}$. The specific configuration of singular dislocation lines
is prescribed by appropriately chosen $e^P_{i j}$ since the
plastic strain is concentrated on cut surfaces bounded by the
dislocation lines \cite{dv, dv1}. The present approach enables one to avoid
the stress divergencies on the dislocation lines. In this respect, an
auxiliary field $C_{i j}$ (which is not a variable of the functional integration) is
postulated to ensure the background fixing at $\ell\ne 0$ in such way that the
components of ${\si}^{\rm b}_{i j}$ just coincide with the conventional
dislocation stresses governed by $e^P_{i j}$ ($C_{i j}$ vanishes at $\ell
= 0$).

The notation $\cD ({\si}^{\rm b}_{i j}, {\si}^{\rm c}_{i j}, u_i,
e_{i j})$ implies the integration measure in (\ref{cor:partf01})
equal to a product of particular measures for each functional
variable marked inside the brackets. More details on the definition of the
functional integration measure by means of the appropriate limiting
procedure can be found in \cite{pop3, pop4, riv, yr, kl3, zinn}. The
normalization factor $1/N$ is to absorb physically irrelevant
multiplicative infinities. The partition function given by (\ref{cor:partf01}),  (\ref{cor:partf02}) naturally generalizes
the partition function of the array of singular dislocations \cite{kl11, kl12}. We shall apply the steepest descent approximation  \cite{pop3, pop4, riv, yr, kl3, zinn} to ${\cZ}$ given by (\ref{cor:partf01}), (\ref{cor:partf02}). The corresponding saddle point solutions are determined from requirement that the variation of the functional $W$ vanishes, $\dl W =0$, under independent variations of its functional arguments.

Firstly, the following observation concerning $\cZ$ (\ref{cor:partf01})
is in order provided three-dimen\-si\-on\-al Eqs.~(\ref{cor:en1}) are
viewed at $\ell=0$. Shifting subsequently
${\si}^{\rm c}_{i j}\,\rightarrow\,{\si}^{\rm c}_{i j} +
{\hat\si}_{i j}$ and $e_{i j}\,\rightarrow\,e_{i j}+{\hat e}_{i
j}$ (here, ${\hat\si}_{i j}$ and ${\hat e}_{i j}$ are new
adjustable fields), one can remove in $E$ the terms
linear in ${\si}^{\rm c}_{i j}$ and $e_{i j}$. Then, the
integrations over ${\si}^{\rm c}_{i j}$ and $e_{i j}$ are
decoupled and can be compensated by redefinition of $1/N$.
Eventually, $E$ is reduced to $E_{\rm el}=E_{\rm el}({\si}^{\rm
b}_{i j})$, and two remaining integrations, over $u$ and
${\si}^{\rm b}_{i j}$, result
in the partition function of the array of singular dislocations
characterized by a specific choice of $e^P_{i j}$ \cite{kl11, kl12}.

Our framework is that of the plane elasticity since it is assumed that the cylinder is long enough \cite{teod, hirth}.
The independence on the third coordinate reduces our study to the two dimensional problem. Therefore, the functionals (\ref{cor:en1}) take the form:
\begin{equation}
\begin{array}{rcl}
&&E_{\rm el}\,=\,\displaystyle{\frac{1}{2\mu}\int
({\si}^{\rm b}_{i} + {\si}^{\rm c}_{i})^2\,\d^2x}\,,\\
&&E_{\rm core}\,=\,2 \displaystyle{\int
(\ell\,e_{i}({\text{inc}}\,e)_{i}\,-\,e_{i}\,
{\si}^{\rm c}_{i})\,\d^2x }\,, \\
&&E_{\rm ext}\,=\,\displaystyle{\int{\si}^{\rm b}_{i} (\cd_i
u - 2\cP_{i})\,\d^2x}\,,
\end{array}
 \label{cor:en2}
\end{equation}
where the integrands are $(x_1, x_2)$-dependent, and summation
goes over $i=1, 2$. Since the displacement vector of the straight
screw dislocation is along $Ox_3$ \cite{teod}, we use $u\equiv
u_3$. In addition, we abbreviate: ${\si}^{\#}_{i}\equiv
{\si}^{\#}_{i 3}$ ($\#$ is $\rm b$ or $\rm c$), ${e}_{i}\equiv
{e}_{i 3}$, etc.
For instance, the notation for the functional integration measure is of the form: $\cD ({\si}^{\rm b}_{i}, {\si}^{\rm c}_{i}, u,
e_{i})$.

Let us consider equations (\ref{cor:en2}) at $\ell=0$. The source
$\cP_{i}$ is taken at $\ell=0$ in the form: $\cP_{i}=e^P_{i}$. A
single positively oriented straight screw dislocation intersecting
the plane $x_1 O x_2$ at ${\bf x}={\bf y}$ (we use ${\bf x}\equiv
(x_1, x_2)$) is ``produced'' by the plastic strain with a single
non-zero component $e^P_{2}=(b/2) \theta(-x_1+y_1) \dl(y_2-x_2)$
\cite{dv, dv1}. Here, $\theta(\cdot)$ is the Heaviside function and $b$
is absolute value of the Burgers vector ${\bf b}$ lying along
$Ox_3$. Furthermore, the integration over $u$ is equivalent to the
insertion of the delta-functional $\dl(\cd_i {\si}^{\rm
b}_{i})$. The equilibrium equation $\cd_i {\si}^{\rm
b}_{i}=0$ is fulfilled by the Kr\"oner ansatz ${\si}^{\rm
b}_{i}\equiv \mu\,\epsilon_{i j}\,\cd_j f^{\rm b}$ (see
\cite{lh}), where
$\epsilon_{i j}$ is totally antisymmetric symbol of second rank. Vanishing of the variation $\dl W$ under the shift
${f}^{\rm b} \rightarrow {f}^{\rm b} + \,\dl{f}^{\rm b}$ allows one to
determine the potential ${f}^{\rm b}$.

Consider a collection of $\cN$ screw dislocations with the coordinates
${\bx}={\bf y}_I$ and the Burgers
vectors ${\bf b}_I$, $1\le I\le \cN$. A disc of radius $R$ is a cross-section of the cylinder containing the dislocations. Assume that the defect positions on the disc are well separated from each other as well as from the boundary (the distribution of the dislocations is not too dense). For our purposes it is appropriate to consider the solution corresponding to the infinite plane, i.e., the influence of the boundary is neglected at a large enough $R$.
Taking into account arbitrariness of $\dl{f}^{\rm b}$, we obtain the extremum condition:
\begin{equation}
\Dl{f}^{\rm b}\,=\,2 i (\cd_1 e^P_{2}\,-\,\cd_2
e^P_{1})\,=\,- i \rho({\bf x})\,,\qquad \rho({\bf x})\,\equiv\,\sum\limits_{I=1}^{\cN} j({\bf x}-{\bf y}_I)\,, \label{cor:eq1}
\end{equation}
where $j({\bf x}-{\bf y}_I)\equiv b_I {\stackrel{(2)} \dl}({\bf
x}-{\bf y}_I)$, $\stackrel{(2)}{\dl}\!({\bx})$ is the
delta-function on $x_1Ox_2$, and $\rho({\bf x})$ is thus the dislocation density.
We determine ${f}^{\rm b}$ from
(\ref{cor:eq1})
using the Prandtl stress potential \cite{lh} for the collection of the screw
dislocations:
\begin{equation}
{f}^{\rm b}\,=\,\frac{-i}{2\pi} \sum\limits_{I=1}^{\cN} b_I \log |{\bx}-{\bf y}_I| \,. \label{cor:eq2}
\end{equation}

Now we turn to $\ell\ne 0$ and require vanishing of $\dl W$
for independent variations of the functional arguments of $W$. Generally, provided that Eqs.~(\ref{cor:en1}) are considered, the
variation of ${\si}^{\rm c}_{i j}$ results in the `strain--stress'
constitutive law:
\begin{equation}
e_{i j}\,=\,\frac{1}{2\mu}\Bigl({\si}^{\rm b}_{i j} + {\si}^{\rm
c}_{i j} - \frac{\nu}{1+\nu}\bigl({\si}^{\rm b}_{l l} + {\si}^{\rm
c}_{l l}\bigr)\dl_{i j}\Bigr)\,.
 \label{cor:en11}
\end{equation}
Also, the variation of $e_{i j}$ gives the equation:
\begin{equation}
({\text{inc}}\,e)_{i j}\,=\,\frac{1}{2 \ell}\,{\si}^{\rm c}_{i
j}\,, \label{cor:eq4}
\end{equation}
which is a linearized version of the Einstein-type gauge equation investigated in the non-linear approach of \cite{mal2}. Since $\cd_i({\text{inc}}\,e)_{i j}\equiv 0$, we put (in the plane problem) ${\si}^{\rm
c}_{i}=\mu \,\epsilon_{i j}\,\cd_j {f}^{\rm c}$ to fulfil
$\cd_i {\si}^{\rm c}_{i} = 0$. Arbitrariness of the variations
$\dl{f}^{\rm b}$ and $\dl{f}^{\rm c}$ results in the extremum conditions:
\begin{equation}
\begin{array}{l}
\Dl\bigl({f}^{\rm b}+{f}^{\rm c}\bigr)\,=\, \displaystyle{2 i
(\cd_1 \cP_{2}\,-\,\cd_2
\cP_{1})}\,,\\[0.3cm]
\Dl\bigl({f}^{\rm b}+{f}^{\rm c}\bigr)\,=\,\kappa^2 {f}^{\rm
c}\,,\qquad \displaystyle{\kappa^2\equiv\frac\mu \ell}\,,
\end{array}
\label{cor:eq5}
\end{equation}
where $\cP_{i}\equiv e^P_i+C_i$, and Eqs.~(\ref{cor:eq1})--(\ref{cor:eq4}) are accounted for.
Self-consistency of Eqs.~(\ref{cor:eq5}) requires coincidence of their right-hand sides and takes the form:
\begin{equation}
\cd_1 C_{2}\,-\,\cd_2 C_{1}\,=\,
\frac12\,(\rho - i \kappa^2{f}^{\rm
c})\,,
\label{cor:eq50}
\end{equation}
where $\rho\equiv\rho({\bf x})$ is defined in (\ref{cor:eq1})
We represent a general solution to
(\ref{cor:eq50}) as follows:
\begin{equation}
C_i\,=\,
\displaystyle{\frac{i}{2}} (\cd_i \phi\,+\,\epsilon_{i j}\,\cd_j \psi)\,,
\label{cor:eq52}
\end{equation}
where $\phi$ is the arbitrary regular function, while $\psi$ is to be
found after substitution of (\ref{cor:eq52}) to (\ref{cor:eq50}).
If (\ref{cor:eq50}) is respected, either ${f}^{\rm b}$ or
${f}^{\rm c}$ can be fixed arbitrarily. Our strategy at $\ell\ne 0$ is to keep
${f}^{\rm b}$ still respecting Eq.~(\ref{cor:eq1}).
Then Eqs. (\ref{cor:eq5}) lead to a single equation for~${f}^{\rm
c}$:
\begin{equation}
(\Dl\,-\,\kappa^2) {f}^{\rm c}\,=\,i \rho({\bf x}) \,, \label{cor:eq51}
\end{equation}
which is solved as follows:
\begin{equation}
{f}^{\rm c}\,=\,\frac{-i}{2\pi}\,
\sum\limits_{I=1}^{\cN} b_I K_0(\kappa|{\bx}-{\bf y}_I|)
\,, \label{cor:eq6}
\end{equation}
where $K_0(\cdot)$ is the modified Bessel function. The
choice $\phi=0$, $\psi={f}^{\rm c}$ ensures coincidence of
(\ref{cor:eq50}) and (\ref{cor:eq51}).

Therefore, $\frac{1}{i}({f}^{\rm b}+{f}^{\rm c})$ is the stress potential of the collection of non-singular screw
dislocations. The fields ${\si}^{\rm b}_{i}$ and
${\si}^{\rm c}_{i}$ are rather the saddle point solutions of the
functional $W$ (\ref{cor:partf02}), while the proper stress fields are given by $\frac{1}{i} {\si}^{\rm b}_{i}$ and
$\frac{1}{i} {\si}^{\rm c}_{i}$. Equations (\ref{cor:eq2}) and
(\ref{cor:eq6}) demonstrate that a modification of the Prandtl stress potential $\frac{1}{i}{f}^{\rm b}$ occurs within the
cores, i.e., within the tubular vicinities of transverse
size $\simeq \kappa^{\1}$. When $\kappa
|{\bx}-{\bf y}_{I}|\gg 1$, $\forall I$, the total solution is dominated by
the conventional contribution $\frac{1}{i} {f}^{\rm b}$. The sum $\frac{1}{i}({f}^{\rm b}+{f}^{\rm c})$ behaves smoothly when ${\bx}$ is close to either of ${\bf y}_{I}$, and the known singularities in the
total stress distribution $\frac{1}{i} ({\si}^{\rm
b}_{i}+{\si}^{\rm c}_{i})$ do not appear.

Thus, the stationary solution corresponds to the modified
screw dislocation \cite{mal1} which is, in turn, in agreement with the solution obtained by means
of the gradient elasticity \cite{g1, g2}. The gradient elasticity
itself belongs to a class of the generalized continuum theories
which effectively take into account interatomic forces for the
explanation of the material behaviour on the nano-scales (and thus inside
the defect cores) \cite{pss}. The solution obtained will be used to estimate the partition function given by
(\ref{cor:partf01}), (\ref{cor:partf02}) and (\ref{cor:en2}) using
the steepest descent approximation.

Let us proceed with the consideration of the partition function $\cZ$.
Integration by
parts in $W$ transforms $e_{i} ({\text{inc}}\,e)_{i}$ into the
quadratic expression, $-{\frac{1}2} (\cd_i e_{j}-\cd_j e_{i})^2$, invariant under the shift $e_{i} \rightarrow e_{i} + \cd_i
g$, where $g$ is arbitrary function (the Abelian gauge
transformation). To restrict the gauge arbitrariness related to the functional
integration over $e_{i}$, let us impose, so-called, ``Coulomb gauge'' $\cd_i e_{i}=0$ \cite{pop3}. Finally, the contribution
$\ell e_{i} ({\text{inc}}\,e)_{i}$ in $W$ is replaced by $\ell
e_{i} \Dl e_{i}$, and the shift
\begin{equation}
e_{i}\,\rightarrow\,e_{i}\,+\,(2 \ell)^{-1}\,\Dl^{-1}{\si}^{\rm
c}_{i} \label{cor:repl1}
\end{equation}
cancels the term linear in $e_{i}$. The
Green function of two-dimensional Laplacian $\Dl^{-1}$ acts as convolution operator. The resulting
Gaussian integration over $e_{i}$,
\[
\int e^{-2 \be \ell \int e_{i}\Dl e_{i}\,\d^2 x}\,\cD(e_{i})\,,
\]
is absorbed into $1/N$, and the partition function $\cZ$ takes the form:
\begin{eqnarray}
&&{\cZ}\,=\,\frac1N\int e^{-\be{\w W}}\cD ({\si}^{\rm b}_{i},
{\si}^{\rm c}_{i}, u)\,,
\label{cor:partf11}\\
&&\w W\,\equiv\,{\w E}\,-\,i E_{\rm ext}\,,\quad
 {\w E}\equiv E_{\rm el}+{\w E}_{\rm core}
\,,
 \label{cor:partf12}
\end{eqnarray}
where $E_{\rm ext}$ is given by (\ref{cor:en2}), and
${\w E}_{\rm core}$ is expressed as follows:
\begin{equation}
\displaystyle{{\w E}_{\rm
core}}\,\equiv\,\displaystyle{\frac{-1}{2 \ell}\int {\si}^{\rm
c}_{i}\,\Dl^{-1}{\si}^{\rm c}_{i} \,\d^2x}\,. \label{cor:partf2}
\end{equation}
The functional $\w W$ includes the elastic energy of
non-singular dislocations $E_{\rm el}$, while the core energies are given by (\ref{cor:partf2}). The representations expressed either by
(\ref{cor:partf01}), (\ref{cor:partf02}), (\ref{cor:en2}), or by
(\ref{cor:partf11}), (\ref{cor:partf12}), (\ref{cor:partf2}) are equivalent, i.e., the
same Eqs.~(\ref{cor:eq1}) and (\ref{cor:eq5}) arise for ${f}^{\rm
b}$ and ${f}^{\rm c}$, respectively.

We are mainly interested in the exponential that contributes to the approximate expression for $\cZ$ (\ref{cor:partf11}): $\cZ\simeq ({\Det{\rm A}})^{\1} e^{-\be\CW}$, where ${\rm A}$ is a kernel of the integral operator related to the second variation of $\widetilde W$ \cite{yr}. To derive $\CW$, it is necessary to substitute the solutions $f^{\rm b}$, $f^{\rm c}$ into $\w W$ (\ref{cor:partf12}) and apply the Green theorem. We neglect the terms exponentially small at $\kappa R\gg 1$, as well as those contributions which are irrelevant provided the differences $|{\bf y}_{I J}|\equiv |{\bf y}_I-{\bf y}_J|$ are moderate in comparison with $R$. The
``electro-neutrality'' condition $\sum_{I=1}^{\cN} b_I =0$ prevents divergency caused by a large
$\log R$. Finally, the effective energy ${\CW}$ takes the form in the case of $\cN$ dislocations:
\begin{equation}{\CW}\,=\,\displaystyle{2 \mu
\int\cd_1 e^P_{2}({\bf x})\,
\big((\Dl-\kappa^2)^{\1}\,-\,\Dl^{\1}\big)\,\cd_1 e^P_{2}({\bf s})\,\d^2x\,\d^2s}\,,
 \label{cor:eq14}
\end{equation}
where the plastic sources are expressed like in (\ref{cor:eq1}), i.e.,
$\cd_1 e^P_{2}=-\rho/2$.
Transforming ${\CW}$ (\ref{cor:eq14}) further,
we represent it as the potential dependent on the
dislocation positions $\{{\bf y}_I\}\equiv\{{\bf y}_I\}_{1\le
I\le\cN}$:
\begin{equation}
\begin{array}{rcl}
{\CW}\!& = &\! {\CW}\bigl(\{{\bf y}_I\}
\bigr)=
 \\ [0.3cm]
&=&\!\displaystyle{\frac{-\mu}{4 \pi} \int\rho({\bf x})\,\bigl[\log|{\bf x}-{\bf s}|\,+\,K_0(\kappa |{\bf x}-{\bf s}|)\bigr]\,\rho({\bf s})\,{\rm d}^2x\,{\rm d}^2 s}
\\ [0.4cm]
  &=&\!\displaystyle{\frac{-\mu}{4 \pi}\Bigl(\sum\limits_{I \ne
J}\,b_I b_J\,\bigl [ \log|{\bf
y}_I-{\bf y}_J|\,+\,K_0(\kappa |{\bf
y}_I-{\bf y}_J|)\bigr]+\log\Bigl(\frac{2}{\kappa\ga}\Bigr)\sum\limits_{I}\,b^2_I
\Bigr)}\,,
\\ [0.4cm]
  &=&\!\displaystyle{\frac{-\mu}{4 \pi}\sum\limits_{I \ne
J}\,b_I b_J\,{\CU}(\kappa |{\bf
y}_I-{\bf y}_J|)}\,,
\end{array}
\label{cor:eq15}
\end{equation}
where ${\CU}(\cdot)$ is defined as
\begin{equation}
{\CU}(s)\equiv \log\bigl(\frac\ga 2 s\bigr)\,+\,K_0(s)\,,
\label{cor:eq11}
\end{equation}
and $\log(\frac{2}{\kappa\ga})$ is the core energy of the dislocation in the Coulomb gas representation. The distances in
(\ref{cor:eq15}) are referred, for simplicity, to unit lattice spacing of a cubic crystal.

We turn to calculation of the force $F_{_{I\!
J}}= -\frac{\d{\CW}}{\d | {\bf y}_{_{I\!
J}}|}$. Its expression,
\begin{equation}
\frac{F_{_{I\!
J}}}{L}\,=\,\frac{\mu b_I b_J \kappa}{2\pi}\,f(\kappa|{\bf y}_{I\! J}|)\,,\qquad
f(s)\,\equiv\,\frac{\d{\CU}}{\d s}\,=\,s^{-1}\,-\,K_1(s)\,,
\label{cor:eq111}
\end{equation}
demonstrates that ${\CW}$ (\ref{cor:eq15}) at large separation $|{\bf y}_{I\!J}|$ corresponds to the energy of the Coulomb attraction between
two charges of unlike signs ($b_I b_J < 0$).
The maximal attraction of two opposite dislocations occurs at $|{\bf y}_{I}-{\bf y}_{J}|\approx \frac{1.1}{\kappa}$.

Consider a thermodynamical ensemble of positive and negative modified screw dislocations located, respectively, at $\{{\bf y}^+_I\}_{1\le I\le\cN}$ and $\{{\bf y}^-_I\}_{1\le I\le\cN}$ and possessing unit Burgers vectors. By analogy with the case of the two-dimensional electro-neutral plasma of positive and negative charges \cite{deut}, the corresponding grand-canonical partition function ${\bf Z}_{C}$ is written, with regard at (\ref{cor:eq15}), as follows:
\begin{equation}
\begin{array}{rcl}
{\bf Z}_{C}\!&=&\!\displaystyle{
\sum\limits_{{\cN}=0}^\infty\,\,
\frac{1}{{\cN}!\,{\cN}!} \prod\limits_{I=1}^{\cN} \int\d^2{\bf
y}^+_{I}\,\prod\limits_{J=1}^{\cN} \int\d^2{\bf y}^-_{J}}\,\exp\Bigl[-2 \be{\cN}\La\,+\,
\\[0.5cm]
&+&\!\displaystyle{
\frac{\be\mu}{4\pi}\,\Bigl(\sum\limits_{I \ne
J}\,{\CU}(\kappa |{\bf y}^+_I-{\bf y}^+_J|)\,+\,
\sum\limits_{I \ne
J}\,{\CU}(\kappa |{\bf y}^-_I-{\bf y}^-_J|)
\,-\,2\,\sum\limits_{I \ne
J}\,{\CU}(\kappa |{\bf y}^+_I-{\bf y}^-_J|)
\Bigr)\Bigr]}\,. \end{array} \label{cor:partf6}
\end{equation}
Here, $2 \cN$ is the number of dislocations, and $\La$ is the chemical potential
per dislocation. The representation
(\ref{cor:partf6}) implies that the summation over possible positions of the dislocations is replaced by the integration. The potential ${\CU}(s)$ (\ref{cor:eq11}) is present in (\ref{cor:partf6}) instead of the regularised logarithm used in \cite{ab, ab1} for the case of the
world-sheet vortices. It is
appropriate to recall that the Coulomb gas of point charges
is related to the two-dimensional \textit{sine-Gordon} field theory
\cite{colem, sam, frohl, mond}.

\section{The correlation functions}
\label{sec3}

\subsection{Field-theoretical derivation of the stress-stress correlati\-on\break fu\-n\-ction}

Gaining the experience of the derivation of the partition function
(\ref{cor:partf6}), we pass on to the
investigation of the relevant correlation functions. Let us take the
field theory presented in section 2 as a starting point. Define the two-point
\textit{stress-stress} \textit{correlation} \textit{functions}
$\l{\si}^{\#}_{i}({\bf x}_1)\,{\si}^{\#}_{j}({\bf x}_2)\r_{_P}$ ($\#$
is either $\rm b$ or $\rm c$) by means of the fol\-lowing
functional averages \cite{pop3, pop4, riv, yr, kl3, zinn}:
\begin{equation}
\l{\si}^{\#}_{i}({\bf x}_1)\,{\si}^{\#}_{j}({\bf x}_2)\r_{_P}\,\equiv\,
\frac{1}{\cZ_o}\int {\si}^{\#}_{i}({\bf x}_1)\,{\si}^{\#}_{j}({\bf
x}_2)\, e^{-\be W} \cD ({\si}^{\rm b}_{i}, {\si}^{\rm c}_{i}, u,
e_{i})\,, \label{cor:genf1}
\end{equation}
where the functional $W$ is expressed by
(\ref{cor:partf02}) and (\ref{cor:en2}), and the index $P$ points out
that the correlators (\ref{cor:genf1})
are defined with respect to a prescribed dislocation distribution specified in terms of the plastic strain $e^P_{i}$ ($i=1, 2$). The latter is accounted for through the dependence of the functional $W=W({\cP}_i)$  on the source ${\cP}_i$. In turn, the normalization factor ${\cZ}_o$ is given by (\ref{cor:partf01}) though with $W$
taken at $\cP_{i}=0$ and $\ell=0$. In other words, ${\cZ}_o$ is the partiton function of a defectless state.

Approach of the generating functional is a natural way to evaluate (\ref{cor:genf1}) in the framework of the functional integration. We
introduce the generating functional ${\cG}[ {{\bf J}^{\rm b}},
{{\bf J}^{\rm c}}\,|\,\bcP]$ as the functional integral parametrized
by two auxiliary sources ${{\bf J}^{\rm
b}}$ and ${{\bf J}^{\rm c}}$, as well as by the source $\bcP$:
\begin{equation}
{\cG}[ {{\bf J}^{\rm b}}, {{\bf J}^{\rm c}}\,|\,\bcP]\,=\,\int
e^{-\be W + i \be \int ({J}^{\rm b}_{i} {\si}^{\rm b}_{i}+ {J}^{\rm
c}_{j}{\si}^{\rm c}_{j})\,{\d}^2x}\,\cD ({\si}^{\rm b}_{i},
{\si}^{\rm c}_{i}, u, e_{i})\,. \label{cor:genf2}
\end{equation}
Here the normalization factor is included into the integration
measure, and each bold-faced notation, ${{\bf J}^{\#}}$ ($\#$ is
$\rm b$ or $\rm c$) or $\bcP$, implies two components, say,
${J}^{\#}_1$ and ${J}^{\#}_2$. The source $\bcP$ is standing
separately in left-hand side of (\ref{cor:genf2}) since its
meaning is different. The correlators we are interested in,
$\l{\si}^{\#}_{i}({\bf x}_1)\,{\si}^{\#}_{j}({\bf x}_2)\r_{_P}$,
arise as follows:
\begin{equation}
\l{\si}^{\#}_{i}({\bf x}_1)\,{\si}^{\#}_{j}({\bf
x}_2)\r_{_P}\,=\,\lim_{\substack{{\bf J}^{\rm b},\,{\bf J}^{\rm c}\to
0}}\,\Bigl({\cG}^{\1}[{\bf J}^{\rm b}, {\bf J}^{\rm
c}\,|\,{\bcP}^{\rm ph}_o]\, \frac{(-i/\be)^{2}\,\dl^2}{\dl J^{\#}_{i}({\bf
x}_1)\dl J^{\#}_{j}({\bf x}_2)}\,{\cG}[{\bf J}^{\rm b}, {\bf
J}^{\rm c}\,|\,{\bcP}^{\rm ph}] \Bigr)\,, \label{cor:genf3}
\end{equation}
where ${\bcP}^{\rm ph}$ is substituted for ${\bcP}\,(={\bf e}^P+{\bf C})$ in order to express that ${\cG}$ is taken for a definite parametrization of the background configuration in terms of the field ${\bf C}$ (the appendix provides the calculation of ${\cG}[ {{\bf J}^{\rm b}}, {{\bf J}^{\rm c}}\,|\,{\bcP}^{\rm ph}]$). Moreover, ${\bcP}^{\rm ph}_o \equiv {\bcP}^{\rm ph}\bigl|_{{\bf e}^P=0}$, i.e., the normalization is taken with respect of a defectless state, as in (\ref{cor:genf1}).

We shall obtain the correlation function
$\l{\si}^{\rm tot}_{i}({\bf x}_1)\,{\si}^{\rm tot}_{j}({\bf
x}_2)\r_{_P}$ of the physical stress field ${\si}^{\rm tot}_{i}({\bf
x}) \equiv \frac{1}{i}({\si}^{\rm b}_{i}({\bf x})+{ \si}^{\rm
c}_{i}({\bf x}))$. First, we calculate the functional (\ref{cor:genf2}) by shifts of the integration variables which cancel
the terms linear in ${\bf J}^{\rm b}$, ${\bf J}^{\rm c}$. It is appropriate to obtain the answer in terms of a single source
${J}_{i}({\bf x})={J}^{\rm b}_{i}({\bf x})={J}^{\rm c}_{i}({\bf
x})$. The result is of the form:
\begin{equation}
\displaystyle{{\cG}[{{\bf J}}, {\bf J}\,|\,{\bcP}^{\rm ph}]\,=\,
e^{-\mu\be Q/2 }\,{\cG}[0, 0\,|\,{\bcP}_o^{\rm
ph}] }\,, \label{cor:genf5}
\end{equation}
where
\begin{equation}
\begin{array}{l}
\displaystyle{Q\,\equiv\,\int\left( \cd_i\cJ_i \Big(\frac{1}{\Dl}\,-\,\frac{1}{\Dl-\kappa^2}\Big)
\cd_k\cJ_k\,-\,\cJ_i \frac{\kappa^2}{\Dl-\kappa^2}
\cJ_i\,+\,\Dl {J}_{i}\frac{1}{\Dl-\kappa^2}
{J}_{i}\right) \d^2x}\,,\\[0.5cm]
\cJ_i\equiv{J}_{i}- 2e^P_{i}
\end{array}
\label{cor:genf6}
\end{equation}
(the Green functions $\Dl^{\1}$, $(\Dl-\kappa^2)^{\1}$ imply the convolution operators). Then the physical correlation functions arise as follows:
\begin{equation}
\l{\si}^{\rm tot}_{i}({\bf x}_1) {\si}^{\rm tot}_{j}({\bf
x}_2)\r_{_P}\,=\,\lim_{\substack{{\bf J}\to
0}}\,\Bigl({\cG}^{\1}[{\bf J}, {\bf J}\,|\,{\bcP}^{\rm ph}_o]\,
\frac{\be^{-2}\,\dl^2}{\dl J_{i}({\bf x}_1)\dl J_{j}({\bf
x}_2)}\,{\cG}[{\bf J}, {\bf J}\,|\,{\bcP}^{\rm ph}] \Bigr)\,.
\label{cor:genf4}
\end{equation}
We calculate (\ref{cor:genf4}) using (\ref{cor:genf5}) and
(\ref{cor:genf6}), and obtain:
\begin{equation}
{\l}{\si}^{\rm tot}_{i}({\bf x}_1) {\si}^{\rm
tot}_{j}({\bf x}_2){\r}_{_P}\,=\,\displaystyle{\frac{\mu^2}{4}\,e^{-\be \CW}}\,\lim_{\substack{{\bf J}\to 0}}\Bigl[\bigl(\dl_{J_{i}({\bf x}_1)}
Q\bigr)\bigl(\dl_{J_{j}({\bf x}_2)} Q\bigr)
\displaystyle{-\,
\frac{2}{\mu\be}}\,\dl_{J_{i}({\bf x}_1)}\dl_{J_{j}({\bf x}_2)}Q
\Bigr]\,,
\label{cor:genf7}
\end{equation}
where $\dl_{J_{i}({\bf x})}\equiv \dl/\dl J_{i}({\bf x})$, and $\CW$, $Q$ are given by (\ref{cor:eq15}) and (\ref{cor:genf6}). Equation (\ref{cor:genf7}) is re-expressed after the variational differentiations:
\begin{equation}
\displaystyle{\l{\si}^{\rm tot}_{i}({\bf x}_1) {\si}^{\rm
tot}_{j}({\bf x}_2)\r_{_P}\,=\,e^{-\be\CW}\,\Bigl[
{\si}^{\rm tot}_{i}({\bf x}_1) {\si}^{\rm tot}_{j}({\bf x}_2)
\,-\,\frac{\mu}{2\pi \be}\cd_{({\bf x}_1)_i}\cd_{({\bf x}_2)_j} \CU(\kappa|{\bf x}_1-{\bf x}_2|)\Bigr]} \,,
\label{cor:genf8}
\end{equation}
where
\begin{equation}
\displaystyle{{\si}^{\rm tot}_{i}({\bf x})\,=\,\frac{\mu}{\pi}
\epsilon_{i k}\,\cd_{({\bf x})_k}\int \CU(\kappa|{\bf x}-{\bf
s}|)\,\cd_1 e^P_2({\bf s}) \d{\bf s}} \label{cor:genf9}
\end{equation}
is the total elastic stress of the dislocational configuration. The exponential factor in (\ref{cor:genf8}) is the Boltzmann weight containing in its exponent the energy of the array of the dislocations.

Consider, for a comparison, our theory without the influence of
the core energy, i.e., at $\ell=0$. Derive, firstly, the
stress-stress correlator $\l{\si}^{\rm b}_{i}({\bf x}_1)
{\si}^{\rm b}_{j}({\bf x}_2)\r$ in the absence of the defects. The source $e^P_i$ being treated as unphysical one (that becomes zero after the variation) enables us to obtain:
\begin{equation}
\displaystyle{\l{\si}^{\rm b}_{i}({\bf x}_1) {\si}^{\rm
b}_{j}({\bf x}_2)\r\,=\,\frac{-\mu}{2\pi \be}\,
\displaystyle{\cd_{({\bf x}_1)_i}\cd_{({\bf x}_2)_j} \log|{\bf
x}_1-{\bf x}_2| }} \,. \label{cor:genf10}
\end{equation}
Let the plastic source be the physical one. Then we obtain:
\begin{equation}
\begin{array}{l}
\displaystyle{\l{\si}^{\rm b}_{i}({\bf x}_1) {\si}^{\rm
b}_{j}({\bf x}_2)\r_{_P}}\,=\,e^{-\be\CW}
\Bigl[\displaystyle{\frac{- \mu}{2\pi \be}\cd_{({\bf
x}_1)_i}\cd_{({\bf x}_2)_j} \log|{\bf x}_1-{\bf x}_2|
} \\[0.3cm]
+\,\displaystyle{\frac{\mu^2}{4\pi^2} \sum\limits_{I, J} b_I b_J
\bigl(\epsilon_{i k} \cd_{({\bf x}_1)_k}\log|{\bf x}_1-{\bf y}_I|\bigr)\,\bigl(\epsilon_{j l} \cd_{({\bf x}_2)_l} \log|{\bf x}_2-{\bf y}_J|\bigr) \Bigr]} \,,
\end{array}
\label{cor:genf11}
\end{equation}
where $\CW$ is given by (\ref{cor:eq15}), although with $\CU$
replaced appropriately by the logarithmic potential.

The representation (\ref{cor:genf8}) corresponds to a specific
spatial distribution of collection of the dislocations. We shall average  $\l{\si}^{\rm tot}_{i}({\bf
x}_1) {\si}^{\rm tot}_{j}({\bf x}_2)\r_{_P}$ over positions of
the defects.

\subsection{Mean square of the dipole momentum}

Let us investigate the grand-canonical ensemble of the dislocations using the
so-called \textit{dipole phase} approximation which corresponds to
pairs of the dislocations with opposite signs (located in ${\bf
x}^\pm_I$, $1\le I\le \cN$) bound into ``molecules''. The
corresponding partition function is specialized as follows \cite{ab}:
\begin{equation}
\begin{array}{rcl}
{\bf Z}_{\rm
dip}&=&\displaystyle{\sum\limits_{{\cN}=0}^\infty\,\,
\frac{1}{{\cN}!} \prod\limits_{I=1}^{\cN} \int\d^2{\bxi}_{I}
\int\d^2{\bta}_{I} }
\,\exp\Bigl[-2\be{\cN}\La - \\[0.3cm]
&-&\be\Bigl(\sum\limits_{I=1}^{\cN}
{w}({\bta}_{I})\,+\,\sum\limits_{I\ne J} {w}_{I\!J} \Bigr)\Bigl]\,,
\end{array}
\label{cor:partf7}
\end{equation}
where $w({\bta}_{I})$ is the energy of $I^{\rm th}$ dipole
centered in ${\bxi}_{I}=({\bf x}^+_I+{\bf x}^-_I)/2$ with the
dipole momentum ${\bta}_{I}={\bf x}^+_I-{\bf x}^-_I$, while $w_{I\!J}$ is the energy of interaction between $I^{\rm th}$ and $J^{\rm
th}$ dipoles.

Firstly, we calculate the mean square of the dipole
momentum for a single molecule:
\begin{equation}
\l {\bta}^2 \r\,=\,\displaystyle{\frac{\displaystyle{\int}
\exp(-\be
w({\bta}))\,{\bta}^2\d{\bta}}{\displaystyle{\int}\exp(-\be
w({\bta}))\,\d{\bta}} }\,.\label{cor:eq16}
\end{equation}
The average (\ref{cor:eq16}) has been calculated in \cite{kos2} for electrically-neutral gas of particles with charges
$\pm q$ interacting through the potential:
\begin{equation}
\begin{array}{lrl}
& \displaystyle{2 \La - 2q_I\,q_J
\log\frac{|{\bf x}_I-{\bf
x}_J|}{a}}\,,& \quad |{\bf x}_I-{\bf x}_J| > a\,,\\
& 0\,, & \quad |{\bf x}_I-{\bf x}_J| < a\,,
\end{array}
\nonumber
\end{equation}
where ${\bf x}_I$ is the $I^{\rm th}$ charge position, and
$a$ is the cutoff (e.g., the particle diameter, or
lattice spacing). Moreover, $2\La$ is the energy necessary to
create a pair of opposite charges at the distance $a$. The dipole energy is given by $\be w({\bta}) =
\cK \log(|{\bta}|/a)$, where ${\cK}=2\be q^2$. The average
(\ref{cor:eq16}) takes the form \cite{kos2}:
\begin{equation}
\l {\bta}^2 \r\,=\,\displaystyle{a^2\,\frac{{\cK}-2}{{\cK}-4}\,=
\,a^2\,\frac{{\be} q^2-1}{{\be} q^2-2}}\,,\label{cor:eq17}
\end{equation}
provided that $\displaystyle{\be q^2 >2}$. Therefore, $\l {\bta}^2
\r\approx a^2$ in the limit of zero temperature, while $\l {\bta}^2
\r$ infinitely grows at the temperature $T$ near its critical value $T_c$ given by $\be_c q^2=2$.

Two-particle potential studied in \cite{ab} is expressed as
$\be w({\bta}) = 2 \pi \cK \CU(|{\bta}|)$, where ${\cK}=2\be
q^2$ (here $\be$ is an
effective parameter viewed as inverse temperature), and ${\CU}(\eta)$ is regularised at small $\eta\equiv|{\bta}|$,
${\CU}(\eta)= \frac12\log\frac{a^2+ \eta^2}{a^2}$. We obtain from (\ref{cor:eq16}) at
$\displaystyle{\cK}>2/\pi$:
\begin{equation}
\l {\bta}^2 \r\,=\,\displaystyle{a^2\,\frac{1}{\pi
\cK-2}}\,.\label{cor:eq18}
\end{equation}
The dipole momentum (\ref{cor:eq18}) is strictly smaller than $\l
{\bta}^2 \r$ (\ref{cor:eq17}) at ${\cK}> 4$, i.e., in the second
case the molecules are more compact at small enough temperatures.

The present paper deals with the dipole energy
${\CU}$ (\ref{cor:eq11}), and the average  (\ref{cor:eq16}) is specified as follows:
\begin{equation}
\kappa^2 \l {\bta}^2 \r\,=\,\displaystyle{\,
\frac{\displaystyle{\int_0^\infty \exp(- {\cK} (\log
\eta+K_0(\eta)))\,\eta^3\,\d{\eta}}}{\displaystyle{\int_0^\infty
\exp(- {\cK} (\log \eta+K_0(\eta)))\,{\eta}\,\d{\eta}}}}
\,,\label{cor:eq19}
\end{equation}
where $\cK=\mu b^2\be/2\pi$. The integral in the nominator of
(\ref{cor:eq19}) diverges at $\cK < 4$. Since dissociation is most
probable for pairs of the dislocations with unit Burgers vectors, the dipolar
phase does not exist at the temperature $T > T_c\equiv
\frac{\mu}{8\pi}$.

It has been proposed in \cite{g1, ed1, laz2} to represent the radius of the dislocation core $r_c$
in the form
$r_c\simeq\eta_b/\kappa$, where $\eta_b = 4.0$ \cite{g1}, $\eta_b
= 6.0$ \cite{laz2}, or $\eta_b = 10.0$ \cite{ed1}. In turn, $\kappa^{\1}$ has been selected in terms of interatomic spacing $a$ as follows: $\kappa^{\1} = 0.25\,a$ \cite{g1}, $\kappa^{\1} = 0.399\,a$ \cite{cem1, laz2}. For instance, $r_c
\approx a$ according to \cite{g1}, while $r_c\approx 2.4 a$
according to \cite{cem1, laz2}.
We represent the ratio
(\ref{cor:eq19}) at ${\cK}$ close enough to its ``critical'' value as follows:
\begin{equation}
\l {\bta}^2 \r\,\simeq\,\displaystyle{ \frac{2
\eta_b^2}{\kappa^2(1+2 \eta_b^2 {\sf A}^{\1})
}\,\frac{1}{{\cK}-4}\,,\qquad {\sf A}^{\1}\equiv
\int_0^{\eta_b}\exp (-
4(\log\eta+K_0(\eta)))\,{\eta}\,\d{\eta}}\,,\label{cor:eq20}
\end{equation}
where $\eta_b$ and $\kappa$ are taken according to \cite{cem1}, \cite{g1} or
\cite{laz2}. Moreover, double-sided estimates can be obtained for $\CU(\eta)$ (\ref{cor:eq11}) by adjusting appropriate trial functions. Such estimates enable one to demonstrate that dependence of $\l {\bta}^2 \r$ (\ref{cor:eq19}) on growing ${\cK}$ is characterized by the double inequality:
\begin{equation}
\displaystyle{\frac{1}{{\cK}^2}\,<\,\kappa^2\,\l {\bta}^2 \r\,<\,\frac{1}{{\cK}} }\,.
\label{cor:eq22}
\end{equation}
The potential $\CU(\eta)=\frac 12\log(1+\eta^2)$ can be used in (\ref{cor:eq19}) instead of $\CU$ (\ref{cor:eq11}) for a comparison of extensions of the dipole momenta. According to the estimate (\ref{cor:eq22}), the dipoles are more compact for $\CU$ (\ref{cor:eq11}) at small enough temperatures.

\subsection{The correlator of two stresses}

Now, we turn to the averaging of the
correlation functions (\ref{cor:genf8}) over the dipole positions. We shall follow \cite{ab} where the asymptotic behaviour of the correlation functions of the string models have been
investigated in the presence of the vortex dipoles.  We shall denote the new average as $\lav{\si}^{\rm
tot}_{i}({\bf x}_1)\,{\si}^{\rm tot}_{j}({\bf x}_2)\rav$, and its
expression arises as follows:
\begin{equation}
\begin{array}{rcl}
\lav{\si}^{\rm tot}_{i}({\bf x}_1)\,{\si}^{\rm tot}_{j}({\bf x}_2)\rav&=&\displaystyle{\frac{-\mu}{2\pi \be}\,\cd_{({\bf x}_1)_i}\cd_{({\bf x}_2)_j} \CU(\kappa|{\bf x}_1-{\bf x}_2|)}\\[0.4cm]
&+& \,\,\displaystyle{{\bf Z}_{\rm dip}^{\1}
\!\!\sum\limits_{{\tiny\begin{array}{l} \mathrm{numbers}\,{\rm
of}\, {\rm dipoles},\\{\rm dipole}\,{\rm positions}\end{array}}}
\!\!{\si}^{\rm tot}_{i}({\bf x}_1) {\si}^{\rm tot}_{j}({\bf
x}_2)\,e^{-\be\CW}} \,, \label{cor:dip1}
\end{array}
\end{equation}
where ${\bf Z}_{\rm dip}$ is the partition function
(\ref{cor:partf7}). The dipole positions are confined within a disc of radius $R$.

According to (\ref{cor:eq22}), the dipoles are very compact since
the dipole momenta are not too large at small enough temperature:
$\l {\bta}^2 \r \ll\kappa^{-2}$. Therefore, summation over the
dipole positions can be replaced by integration. We use
the dipole's center of mass and momentum coordinates,
respectively, $\bxi_L=({\bf y}^+_L+{\bf y}^-_L)/2$ and
$\bta_L={\bf y}^+_L-{\bf y}^-_L$ for each pair of opposite
dislocations at ${\bf y}^\pm_L$ forming the $L^{\rm th}$ dipole ($1\le
L\le\cN$). We continue to use the approximation of non-interacting dipoles, and the sum in right-hand side of
(\ref{cor:dip1}) takes the form:
\begin{equation}
\begin{array}{l}
\displaystyle{\sum \limits_{\tiny\begin{array}{l}{{\rm
numbers}\,{\rm ,}\,{\rm positions}}\end{array}}}\,{\si}^{\rm
tot}_{i}({\bf x}_1) {\si}^{\rm tot}_{j}({\bf x}_2)\,
e^{-\be\CW}\,=\\[0.3cm]
=\,\displaystyle{\Bigl(\frac{\mu b}{2\pi}\Bigr)^2\,\epsilon_{i
k} \epsilon_{j l}\,\cd_{({\bf x}_1)_k} \cd_{({\bf x}_2)_l}
\sum\limits_{{\cN}=1}^\infty\,\, \frac{1}{{\cN}!}
\prod\limits_{I=1}^{\cN} \int\d^2{\bxi}_{I} \int\d^2{\bta}_{I}
\,e^{-\be(2\La\,+\,
{w}({\bta}_{I})) }} \\[0.3cm]
\times \sum\limits_{K, L=1}^{\cN}\bigl(\CU(\kappa|{\bf x}_1-{\bf
y}^+_K|)\,-\,\CU(\kappa|{\bf x}_1-{\bf
y}^-_K|)\bigr)
\bigl(\CU(\kappa|{\bf x}_2-{\bf
y}^+_L|)\,-\,\CU(\kappa|{\bf x}_2-{\bf y}^-_L|)\bigr)\,,
\end{array}
\label{cor:dip2}
\end{equation}
where ${\cN}$ is the number of dipoles, and we use for ${\si}^{\rm tot}_{i}({\bf x})$ the representation
(\ref{cor:genf9}). We
estimate $|\bta_L|\ll |{\bf x}-{\bxi}_L|$ (compactness of
the dipoles) and adopt in leading approximation \cite{ab}:
\begin{equation}
\CU(\kappa|{\bf x}-{\bf y}^+_L|)\,-\,\CU(\kappa|{\bf x}-{\bf
y}^-_L|)\,\approx\,-\,(\bta_L, \cd_{{\bf x}})\,\CU(\kappa|{\bf
x}-{\bxi}_L|)\,, \label{cor:dip3}
\end{equation}
where $(\cdot,\cdot)$ stands for the scalar product of
2-vectors. The relation (\ref{cor:dip3}) allows us to
re-express (\ref{cor:dip2}) as follows:
\begin{equation}
\begin{array}{l}
\displaystyle{\Bigl(\frac{\mu b}{2\pi}\Bigr)^2\,\epsilon_{i k} \epsilon_{j l}\,\cd_{({\bf x}_1)_k} \cd_{({\bf x}_2)_l}\,\sum\limits_{{\cN}=1}^\infty\,\, \frac{1}{{\cN}!}
\prod\limits_{I=1}^{\cN} \int\d^2{\bxi}_{I} \int\d^2{\bta}_{I}
\,e^{-\be(2\La\,+\,
{w}({\bta}_{I}))}} \\[0.3cm]
\times \sum\limits_{L=1}^{\cN}(\bta_L, \cd_{{\bf
x}_1})\,\CU(\kappa|{\bf x}_1-{\bxi}_L|)\,(\bta_L, \cd_{{\bf
x}_2})\,\CU(\kappa|{\bf x}_2-{\bxi}_L|)\,.
\end{array}
\label{cor:dip4}
\end{equation}

To proceed with (\ref{cor:dip4}), essential technical task is to
calculate the integral:
\begin{equation}
\displaystyle{\int\d^2{\bxi} \int\d^2{\bta}\,
(\bta, \cd_{{\bf x}_1})\,
\CU(\kappa|{\bf x}_1-{\bxi}|)\,(\bta, \cd_{{\bf x}_2})\,\CU(\kappa|{\bf
x}_2-{\bxi}|)\,e^{-\be(2\La\,+\, {w}({\bta}))} }\,.\label{cor:eq23}
\end{equation}
First, we express the ${\bta}$-integration by means of the
relation
\begin{equation}
\frac{\displaystyle{\int \exp\bigl(-2\be\La - {\cK}
(\log(\frac\ga2\kappa \eta)+K_0(\kappa\eta))\bigr) {\eta}_i}
{\eta}_j\,\d^2{\bta}} {\displaystyle{\int\exp\bigl(-2\be\La -
{\cK} (\log(\frac\ga2\kappa
\eta)+K_0(\kappa\eta))\bigr)\,\d^2{\bta}}}\,=\,\frac{\dl_{i j}}{2}
\l {\bta}^2 \r\,, \label{cor:eq24}
\end{equation}
where ${\cK}=\frac{\mu b^2\be}{2\pi}$, $\eta=|\bta|$, and $\l
{\bta}^2 \r$ is given by (\ref{cor:eq19}). Then, after introducing
the notation $\bar N$ for the dipole density \cite{kos2, ab}:
\begin{equation}
\bar N\,\equiv\,\displaystyle{\int\exp\bigl(-2\be\La - {\cK}
(\log(\frac\ga2\kappa \eta)+K_0(\kappa\eta))\bigr)\,\d^2{\bta}}
\,,\label{cor:eq26}
\end{equation}
the relation (\ref{cor:eq24}) allows us to re-express the integral
(\ref{cor:eq23}) as follows:
\begin{equation}
\displaystyle{\frac{ \l {\bta}^2 \r \bar N}{2}\int\bigl(\cd_{{\bf
x}_1}\, \CU(\kappa|{\bf x}_1-{\bxi}|)\,, \cd_{{\bf
x}_2}\,\CU(\kappa|{\bf
x}_2-{\bxi}|)\bigr)\,\d^2{\bxi}}\,.\label{cor:eq25}
\end{equation}
The Green theorem enables one to carry out the ${\bxi}$-integration in
(\ref{cor:eq25}). Eventually, we use the partition function ${\bf
Z}_{\rm dip}$ (\ref{cor:partf7}) and obtain:
\begin{equation}
\begin{array}{rcl}
&&\displaystyle{
\lav{\si}^{\rm tot}_{i}({\bf x}_1)\,{\si}^{\rm tot}_{j}({\bf x}_2)\rav\,=\,\frac{-\mu}{2\pi \be}\,\cd_{({\bf x}_1)_i}\cd_{({\bf x}_2)_j} \CU(\kappa|\Dl{\bf x}|)}\, \\[0.3cm]
&& \displaystyle{\,-\,\l {\bta}^2 \r \bar N\,\frac{\mu^2
b^2}{4\pi}\,\epsilon_{i k} \epsilon_{j l}\,\cd_{({\bf x}_1)_k}
\cd_{({\bf x}_2)_l}\bigl(\CU(\kappa |\Dl{\bf
x}|)\,}\displaystyle{ +\frac{\kappa |\Dl{\bf x}|}{2}\,K_1(\kappa
|\Dl{\bf x}|)\bigr) } \,,\label{cor:eq27}
\end{array}
\end{equation}
where (and henceforth) $\Dl{\bf x}\equiv {\bf x}_1-{\bf x}_2$.

At large separation of the arguments, $|\Dl{\bf x}|\gg\kappa^{\1}$, the asymptotic of (\ref{cor:eq27}) is
governed by the logarithmic contribution since the modified Bessel
functions decay exponentially.
We use the rule of differentiation of the logarithm,
\begin{equation}
\begin{array}{l}
\cd_{({\bf x}_1)_k} \cd_{({\bf x}_2)_l}\log|{\bf x}_1-{\bf x}_2|\,=\\[0.3cm]
=\,\displaystyle{\frac{1}{|{\bf x}_1-{\bf x}_2|^2}\Bigl(- \dl_{k l}\,+\,2 \frac{({{\bf x}_1}-{{\bf x}_2})_k\,({{\bf x}_1}-{{\bf x}_2})_l}{|{\bf x}_1-{\bf x}_2|^2}\Bigr)\,-\,\pi\,\dl_{k l}\,{\stackrel{(2)} \dl} ({\bx}_1-{\bx}_2)}\,,
\end{array}
\label{cor:eq281}
\end{equation}
as well as the relation $\epsilon_{i k} \epsilon_{j l} = \dl_{i j}
\dl_{k l} - \dl_{i l} \dl_{k j}$, and obtain:
\begin{equation}
\displaystyle{
\lav{\si}^{\rm tot}_{i}({\bf x}_1)\,{\si}^{\rm tot}_{j}({\bf x}_2)\rav}\,=\,\displaystyle{\Bigl(\frac{-\mu}{2\pi \be}\,+\,\bar N \l
{\bta}^2 \r\,\frac{\mu^2 b^2}{4\pi}\Bigr)
\frac{2(\Dl{\bf x})_i\,(\Dl{\bf x})_j- \dl_{i j}|\Dl{\bf x}|^2}{|\Dl{\bf x}|^4}
}+\,{o}(|\Dl{\bf x}|^{-2})\,.
\label{cor:eq28}
\end{equation}
Therefore, the stress-stress correlator (being considered
with dislocations as well as without them) decreases as $|\Dl{\bf x}|^{-2}$ at growing separation $|\Dl{\bf x}|$.

The asymptotic law $|\Dl{\bf x}|^{-2}$ is due to the
analytical structure of the logarithm differentiated. This law
being extrapolated to small distances, $|\Dl{\bf x}|\ll 1$,
would imply a singular behaviour of the stress-stress correlator in
question. In the present approach, the short-distance behaviour of
the correlation function is less singular because of the core influence.
Since the logarithmic contributions are canceled in the potential
$\CU$, the short-distance behaviour is also less singular:
\begin{equation}
\begin{array}{rcl}
 \lav{\si}^{\rm tot}_{i}({\bf x}_1)\,{\si}^{\rm tot}_{j}({\bf
x}_2)\rav& =&\displaystyle{\frac{-{\mu}}{2\pi
\be}\,\frac{{\kappa}^2}{2}\Bigl[-\,\dl_{i
j}\,\Bigl(\frac12\,+\,{\bar N} \l{\bta}^2\r\,\frac{\mu b^2
\be}{4}\,-\,\log\bigl(\frac\ga2 {\kappa}|\Dl {\bf
x}|\bigr)\Bigr)\Bigr.}
\\[0.5cm]
&+&\,\Bigl.\displaystyle{\frac{(\Dl{\bf x})_i\,(\Dl{\bf
x})_j}{|\Dl{\bf x}|^2} \,+\,{\Cal O}(|\Dl
{\bf x}|^2\log|\Dl {\bf x}|)\Bigr]}\,. \label{cor:eq34}
\end{array}
\end{equation}

We considered the approximation of non-interacting dipoles. Dipole-dipole
interaction can also be taken into account in the present framework.
Moreover, second order
corrections to the stress fields inside the core could become
important for sufficiently dense gas of the dipoles \cite{mal2}. However, this should be a subject of a
separate investigation.

\section{The renormalization of the shear modulus}
\label{sec4}

The present section is to investigate the {\it renormalization} of
the shear modulus caused by the modified screw
dislocations.
One should refer to the original paper
\cite{nel12} for the definition of the inverse of the tensor of
re\-nor\-ma\-li\-zed elastic constants in the presence of dislocations. Further details
on the dislocation contribution to the elastic constants can be
found in \cite{nel12} and \cite{rab} (devoted, respectively, to
two- and three-di\-men\-sion\-al situations). Adopting the
relevant definitions \cite{nel12, rab}, we shall consider the
following expression for the renormalized shear modulus $\mu_{\rm
ren}$:
\begin{equation}
\label{cor:ren1} \displaystyle{\frac{1}{ \mu_{\rm
ren}}\,\equiv\,\frac{\be}{\mu^2 \cS}\,\sum\limits_{i, k =1,
2}\,\iint}\lav{\si}^{\rm tot}_{i}({\bf x}_1)\,{\si}^{\rm
tot}_{k}({\bf x}_2)\rav\,\d^2{\bx}_1 \d^2{\bx}_2\,,
\end{equation}
where the correlator is defined by (\ref{cor:dip1}), $\cS$ is the
area of the sample's cross-section. We use (\ref{cor:eq27}) and (\ref{cor:eq281}), and obtain:
\begin{equation}
\label{cor:ren20}
\begin{array}{l}
\displaystyle{\sum\limits_{k =1, 2}\,\lav{\si}^{\rm tot}_{k}({\bf
x}_1)\,{\si}^{\rm tot}_{k}({\bf x}_2)\rav}\,=\, \displaystyle{
\frac{\mu\kappa^2}{2\pi\be}\Bigl(K_0(\kappa |\Dl {\bf x}|)\,+\Bigr.}\\[0.5cm]
+\,\displaystyle{\frac{\al
\mu}{2}\,\kappa |\Dl {\bf x}|\,K_1(\kappa |\Dl {\bf x}|)\Bigr)
\,,\qquad\quad\al\equiv
 \,\frac{\be b^2 \l {\bta}^2\r \bar N }{2}\,.}
\end{array}
\end{equation}
``Non-diagonal'' correlators $\lav{\si}^{\rm tot}_{k}({\bf
x}_1)\,{\si}^{\rm tot}_{l}({\bf x}_2)\rav$, $k\ne l$, are negligible with respect to the two integrations in (\ref{cor:ren1}). The parameter $\al$ (\ref{cor:ren20}) is proportional to the mean area covered by the dipoles, $2\pi \l {\bta}^2 \r \bar N$. Therefore, one obtains from (\ref{cor:ren1}) and (\ref{cor:ren20}) the following answer:
\begin{equation}
\displaystyle{\frac{1}{\mu_{\rm ren}}\,=\,\frac{1}{\mu}\,
{\cC}_1(\kappa R)\,+\,\al \,{\cC}_2(\kappa R)\,,} \label{cor:ren2}
\end{equation}
where the functions ${\cC}_1(\kappa R)$ and ${\cC}_2(\kappa R)$ are given by the modified Bessel functions:
\begin{equation}
\begin{array}{rcl}
\displaystyle{{\cC}_1(\kappa R)}&
=&\displaystyle{1\,-\,2 K_1(\kappa
R) I_1(\kappa R)}\,,
\\[0.4cm]
\displaystyle{{\cC}_2(\kappa R)}
&=& {2\,-\,2 I_1(\kappa R)\bigl(K_1(\kappa
R)\,-\,\kappa R\,K_1^{\prime}(\kappa R)\bigr) }\,,
\label{cor:ren3}
\end{array}
\end{equation}
and $K_1^{\prime}(z)=\frac{\d}{\d z}K_1(z)$. The renormalization rule
(\ref{cor:ren2}) demonstrates the dependence of the shear modulus
$\mu_{\rm ren}$ on the dimensionless parameter $\kappa R$. The
coefficients ${\cC}_1(\kappa R)$ and ${\cC}_2(\kappa R)$ both are
positive and less than unity though tend to unity at $\kappa R\to
\infty$. We obtain the estimates for ${\cC}_1(\kappa R)$,
${\cC}_2(\kappa R)$ at increasing $\kappa R$:
\begin{equation}
\displaystyle{{\cC}_1(\kappa R)\,
\approx\,
1\,-\,\frac{1}{\kappa R}\,+\,\dots\,,\qquad
{\cC}_2(\kappa R)\,
\approx\,
1\,-\,\frac{3}{2 \kappa R}\,+\,\dots}\,,
\label{cor:ren4}
\end{equation}
where the ellipsis imply the terms ${\cal O}((\kappa R)^{-2})$.

On the other hand, an analogue of
(\ref{cor:ren20}) valid for singular screw dislocations takes the form:
\begin{equation}
\label{cor:ren50} \displaystyle{\sum\limits_{k =1,
2}\,\lav{\si}^{\rm b}_{k}({\bf x}_1)\,{\si}^{\rm b}_{k}({\bf
x}_2)\rav\,=\, \frac{\mu}{\be}(1\,+\,\al \mu )\,{\stackrel{(2)}
\dl} ({\bx}_1-{\bx}_2)} \,.
\end{equation}
The right-hand sides of (\ref{cor:ren20}) and (\ref{cor:ren50}) are
weakly coinciding at $\kappa^{\1}\to 0$ (i.e., when the core's
scale is shrunk). This can be demonstrated by means of integration
with an appropriate trial function. Inserting (\ref{cor:ren50})
into (\ref{cor:ren1}) one obtains:
\begin{equation}
\displaystyle{\frac{1}{\mu_{\rm ren}}\,=\,\frac{1}{\mu}\,+\,
\al}\,. \label{cor:ren5}
\end{equation}
The renormalization rule (\ref{cor:ren5}) is in agreement with
that obtained in the original paper \cite{kos2} with the help of the
macroscopic stress function of collection of the dislocation
dipoles. Equation (\ref{cor:ren5}) agrees with the renormalization
of the shear modulus found in \cite{nel12, rab} provided the
concentration of the defects is low. As is seen from
(\ref{cor:ren5}), increasing of $\al$ results in decreasing of
$\mu_{\rm ren}$. Clearly, Eq.~(\ref{cor:ren2}) is reduced to
(\ref{cor:ren5}) when $\kappa R$ tends to infinity, and so the
unit values of ${\cC}_1$ and ${\cC}_2$ correspond to
the case of singular dislocations. Roughly speaking, shrinking up
the core regions one goes back to singular dislocations.

Equation (\ref{cor:ren2}) can be re-expressed as follows:
\begin{equation}\mu\,\rightarrow\,{\mu}_{\rm ren}\,
=\,\frac{\mu}{{\cC}_1(\kappa R)}\Bigl(1+\mu
\al\,\frac{{\cC}_2(\kappa R)}{{\cC}_1(\kappa R)}\Bigr)^{\1}\,.
\label{cor:ren6}
\end{equation}
At finite $\kappa R$ (the scale $1/\kappa$ is comparable with the sample's scale $R$), the non-triviality of the
dislocation cores is valuable for the renormalization
of the shear modulus. According to (\ref{cor:ren4}) and
(\ref{cor:ren6}), the character of the decreasing of $\mu_{\rm
ren}$ is changed in comparison with the case of the singular
dislocations: it is slower since ${\cC}_2(\kappa R)/{\cC}_1(\kappa
R)<1$ at large $\kappa R$.
Recall that the critical exponents of appropriate correlation
functions are analogously renormalized because of the presence of the
vortex pairs, for instance, in the two-dimensional Bose gas \cite{pop2}, in the classical
planar Heisenberg model \cite{jos},
or on the string world-sheets \cite{ab, ab1}.

\section{Discussion}
\label{sec5}

An array of parallel singular screw dislocations is equivalent (as a
planar system with respect of the sample's cross-section) to the
two-dimensional Cou\-lo\-mb gas of charged point particles. The
latter is a subject covered by the theory \cite{ber1, ber2, kos1,
kos2, pop1, pop2, holz, nel1, nel12, nel13, nel2, kl11, kl12, kl2} of the phase
transitions in the low-dimensional systems of condensed matter
physics. Dislocations influence re-normalization of the elastic
constants of the corresponding material sample. This paper
investigates the re-normalization of the shear modulus in the
case of the screw dislocations possessing the core
regions of finite-size. The transformation to the Coulomb-like system is used.

The approach of \cite{mal1, mal2} to description of singularityless dislocations is elaborated in the given paper
further for studying the collection of the modified screw dislocations as a thermodynamic ensemble. Specifically, a long enough cylinder pierced by non-singular screw dislocations is studied. A field-theoretical formalism is
developed for investigating the corresponding partition function in
the form of the functional integral. The stress potentials of the modified dislocations play the role of its saddle points because of the choice of the energy
functional. The plastic external source is involved which governs
the background stress distribution. This is distinct from \cite{ed1, mal1, laz2, laz3} where the usage of the so-called ``null-Lagrangian'' is responsible for arising of the background stress field.

Calculation of the partition function relates the system of dislocation dipoles to equivalent description of the
electrically neutral Coulomb gas of charges interacting {\it via} potential which is logarithmic at large separation but tends
to zero for the charges sufficiently close to each other. The smoothing of the Coulomb potential at short mutual separations occurs since the self-energy of the cores is accounted for. The stress-stress correlation functions are obtained and
used for studying the renormalization of the shear modulus in the approximation of the dilute gas of the dislocation dipoles.

It is demonstrated that the renormalized
shear modulus depends non-conventionally on the ratio $R/\kappa^{\1}=\kappa R$ of two lengths characterizing the sample's cross-section and the dislocation core sizes. Note that applicability of the effects of renormalization of the elastic
constants to experimental observations is discussed in \cite{rab} for the case of singular defects. Since the contributions caused by the core are
sensible at moderate $\kappa R$, it is hopeful that the formalism
developed could be efficient for the nanotubes with comparable $R$ and
$\kappa^{\1}$. With regard to \cite{mal1,
laz3, g2}, it is hopeful that the formalism presented can be extended to a hollow cylinder,
as well as developed further to cover the modified edge dislocations.
The latter could be interesting as far as the physics of
multi-layer nanotubes and wrapped crystals is concerned
\cite{dkl, graph1}.

\section*{Acknowledgement}

It is a pleasure to express my gratitude to N.~M.~Bogoliubov,
M.~Yu.~Gutkin, M.~O.~Ka\-ta\-na\-ev, and A.~G.~Pronko. The
research described has been supported in part by RFBR
(No.~10-01-00600) and by the Russian Academy of
Sciences program ``Mathematical Methods in Non-Linear Dynamics''.

\section*{Appendix}

Consider the generating functional ${\cG}[{\bf J}^{\rm b},
{\bf J}^{\rm c}\,|\,\bcP]$ (\ref{cor:genf2}) dependent on three
2-com\-po\-nent sources, ${\bf J}^{\rm b}$, ${\bf J}^{\rm c}$, and
$\bcP$:
$$
{\cG}[ {{\bf J}^{\rm b}}, {\bf J}^{\rm c}\,|\,{\bcP}]\,=\,\int
e^{-\be W + i \be \int ({J}^{\rm b}_{i} {\si}^{\rm b}_{i}+ {J}^{\rm
c}_{j}{\si}^{\rm c}_{j})\,{\d}^2x}\,\cD ({\si}^{\rm b}_{i},
{\si}^{\rm c}_{i}, u, e_{i})\,. \eqno{(A1)}
$$
The exponent in (A1) is specified, after fixing the ``Coulomb gauge'', as follows:
$$
\begin{array}{rcl}
&&-\be W +  i \be \displaystyle{\int} ({J}^{\rm b}_{i} {\si}^{\rm b}_{i}+ {J}^{\rm c}_{j}{\si}^{\rm c}_{j})\,{\d}^2x \,=\\[0.5cm]
&&=\,\displaystyle{\frac{-\be}{2\mu}\int
\bigl({\si}^{\rm b}_{i} + {\si}^{\rm c}_{i}\bigr)^2\,\d^2x}\,-\,2 \be\ell \displaystyle{\int e_{i}\Dl\,e_{i}\,\d^2x } \\[0.5cm]
&&+\,i \be \displaystyle{\int{\si}^{\rm b}_{i} \bigl(\cd_i u - 2\cP_{i}
+ {J}^{\rm b}_{i}\bigr)\,\d^2x\,+\,\be\int {\si}^{\rm c}_{i}
\bigl(2 e_i + i{J}^{\rm c}_{i}\bigr)\,\d^2x}\,.
\end{array}
 \eqno{(A2)}
$$
It is appropriate to calculate (A1) by shifts of the  functional
integration variables \cite{pop3, pop4, riv, yr, kl3, zinn}.
As a first step, the strain field $e_{i}$ should be integrated out
by the shift
$$
e_{i}\,\longrightarrow\,e_{i}\,+\,\frac{1}{2 \ell \Dl}\,{\si}^{\rm
c}_{i}\,.
$$
After re-arrangements we obtain for ${\cG}[ {{\bf J}^{\rm b}},
{{\bf J}^{\rm c}}\,|\,{\bcP}]$:
$$
\begin{array}{rcl}
&&{\cG}[ {{\bf J}^{\rm b}}, {{\bf J}^{\rm
c}}\,|\,{\bcP}]\,=\,{\rm const}\times\displaystyle{\int} \exp\left[ \,
\displaystyle{\frac{- \be}{2 \mu}\int {\si}^{\rm c}_{i}\,{\sf
D}^{-1}{\si}^{\rm c}_{i} \,\d^2x}\right. \,+\,i \be \int {\si}^{\rm
c}_{i}\,{J}^{\rm c}_{i}\,\d^2x
   \\[0.5cm]
&&\displaystyle{
 -\,\frac{\be}{2\mu}\int}\Bigl(
{\si}^{\rm b}_{i} {\si}^{\rm b}_{i}\,-\,\left.\displaystyle{
i 2 \mu\,{\si}^{\rm b}_{i} \bigl(\cd_i u - 2\cP_{i} +
{J}^{\rm b}_{i} + \frac{i}{\mu} {\si}^{\rm
c}_{i}\bigr)}\Bigr)\,\d^2x \right]\,\cD ({\si}^{\rm b}_{i},
{\si}^{\rm c}_{i}, u)\,.
\end{array}
 \eqno{(A3)}
$$
Decoupled integrations,
being constant factors which are not of interest now, are systematically included into the prefactor $const$. The kernel ${\sf
D}^{-1}$ is defined in (A3) as follows:
$$
{\sf D}^{-1}\,\equiv\,\dl\,-\,
\displaystyle{\frac{\kappa^2}{\Dl}}\,,
\eqno{(A4)}
$$
where ${\Dl}^{\1}$ is the Green function of two-dimensional
Laplacian, and $\dl$ is the delta-function.

Next step is to shift subsequently the variables ${\si}^{\rm
b}_{i}$ and ${\si}^{\rm c}_{i}$. It is assumed that ${J}^{\rm b}_{i}={J}^{\rm
c}_{i}= J_{i}$.
After the shifts
$$
\begin{array}{rcl}
&&\displaystyle{{\si}^{\rm b}_{i} \,\longrightarrow\,{\si}^{\rm b}_{i}\,+\, i \mu\,(\cd_i u - 2 e^P_{i} + {J}_{i})}\,,\\[0.2cm]
&&\displaystyle{{\si}^{\rm c}_{i} \,\longrightarrow\,{\si}^{\rm
c}_{i}\,-\,i \mu\,{\sf D} (\cd_i u - 2 e^P_{i} )}\,,
\end{array}
\eqno{(A5)}
$$
the generating functional takes the form:
$$
\begin{array}{l}
\displaystyle{ {\cG}[ {{\bf J}}, {{\bf
J}}\,|\,\bcP]\,=\,{\rm const}\times\int} \exp\left[\frac{-\mu\be}{2}\int\right. \bigl(
(\cd_i u + {\cJ}_{i}) (\dl+{\sf D}) (\cd_i u + {\cJ}_{i})+ {J}_{i}{\sf D}{J}_{i}\Bigr.  \\[0.6cm]
-\Bigl.2 \displaystyle{(\cd_i u + {\cJ}_{i})}({\sf D}{J}_{i}+2
C_{i}) \bigr)\,\d^2x\,-\,i \be \int{\si}^{\rm b}_{i} \bigl(2
C_{i}-{\sf D} (\cd_i u - 2 e^P_{i}) \bigr)\,\d^2x\Bigr.
\\[0.6cm]
\left.\displaystyle{-\,\frac{\be}{2 \mu}\int ({\si}^{\rm
c}_{i}\,{\sf D}^{-1}{\si}^{\rm c}_{i} + {\si}^{\rm b}_{i}
{\si}^{\rm b}_{i} + 2{\si}^{\rm b}_{i} {\si}^{\rm c}_{i})
\,\d^2x}\right]\, \cD ({\si}^{\rm b}_{i}, {\si}^{\rm c}_{i}, u)
\,,
\end{array}
\eqno{(A6)}
$$
where the operator ${\sf D}$ is inverse to ${\sf D}^{-1}$ (A4), and $\cJ_i\equiv{J}_{i}- 2e^P_{i}$.

The experience of section~2 tells us about a necessity of fixing of the background contribution by means of a self-consistent choice of the functions $C_{i}$. From a viewpoint of the functional integration,
the corresponding substitute
$$
C_{i}\,=\,\frac12\,{\sf D}\,(\cd_i u - 2 e^P_{i} ) \eqno{(A7)}
$$
(which solves (\ref{cor:eq50})) allows one to decouple the integration over $u$. The corresponding functional integral is just responsible for the dependence of the partition
function on the plastic strain $e^P_{i}$ and thus on the defect
distribution. Then the following representation arises:
$$
\begin{array}{rcl}
\displaystyle{ {\cG}[ {{\bf J}}, {{\bf J}}\,|\,{\bcP}^{\rm
ph}]\,=\,{\rm const}\times\int \exp\left[\frac{-\mu\be}{2}\int\bigl( (\cd_i
u + {\cJ}_{i}) (\dl-{\sf D}) (\cd_i u + {\cJ}_{i})+ {J}_{i}{\sf
D}{J}_{i}\bigr) \,\d^2x \right]\cD (u)} \,,
\end{array}
\eqno{(A8)}
$$
where the notation ${\bcP}^{\rm ph}$ implies that the arbitrariness due to $C_i$ is removed. As a final
step, the shift
$$
u\,\longrightarrow\,u\,-\,\frac{1}{\Dl}\,\cd_i {\cJ}_{i}\,,
$$
allows one to get rid of the contribution of the first order in $\cd_i
u$. Finally, the generating functional takes the form:
$$
\begin{array}{rcl}
\displaystyle{ {\cG}[{\bf J}, {\bf J}\,|\,{\bcP}^{\rm
ph}]}&=&\displaystyle{{\cG}[0, 0\,|\,{\bcP}_o^{\rm
ph}]\,\times}\,\exp\left[
\displaystyle{\frac{-\mu\be}{2}} \int\left( {\cd}_i{\cJ}_i \Big(
\frac{1}{\Dl}\,-\,\frac{1}{\Dl-\kappa^2}\Big)
{\cd}_k {\cJ}_k\right.\right. \\[0.7cm]
 &-&\!\cJ_i\!\left.\left.\displaystyle{ \frac{\kappa^2}{\Dl-\kappa^2}
\cJ_i\,+\,\Dl {J}_{i}\frac{1}{\Dl-\kappa^2} {J}_{i} }\right) d^2x
\right]\,,
\end{array}
\eqno{(A9)}
$$
where $\cJ_i\equiv{J}_{i}- 2e^P_{i}$. All the integrations decoupled are gathered within ${\cG}[0, 0\,|\,{\bcP}_o^{\rm
ph}]$, where ${\bcP}_o^{\rm
ph}\equiv {\bcP}^{\rm
ph}\bigl|_{e_i^P=0}$. Equation (A9) is just the answer expressed by (\ref{cor:genf5}) and (\ref{cor:genf6}).


\begin{thebibliography}{99}

\bibitem{ber1}
        V. L. Berezinskii, Zh. Eksp. Teor. Fiz. {\bf 59} (1970),
        907--920
\bibitem{ber2}
        V. L. Berezinskii, Zh. Eksp. Teor. Fiz. {\bf 61} (1971), 1144--1156
\bibitem{kos1}
        J. M. Kosterlitz, D. J. Thouless, J. Phys. C: Solid
        State Phys. {\bf 5} (1972), L124--L126
\bibitem{kos2}
        J. M. Kosterlitz, D. J. Thouless, J. Phys. C: Solid
        State Phys. {\bf 6} (1973), 1181--1203
\bibitem{pop1}
        V. N. Popov, Teor. Mat. Fiz. {\bf 11} (1972), 354--365
\bibitem{pop2}
        V. N. Popov, Zh. Eksp. Teor. Fiz. {\bf 64} (1973), 674--680
\bibitem{holz}
         A.~Holz, J.~T.~N.~Medeiros, Phys. Rev. B {\bf 17} (1978), 1161--1174
\bibitem{nel1}
         D.~R.~Nelson, Phys. Rev. B {\bf 18} (1978), 2318--2338
\bibitem{nel12}
         D.~R.~Nelson, B.~I.~Halperin, Phys. Rev. B {\bf 19} (1979), 2457--2484
\bibitem{nel13}
         A. P. Young, Phys. Rev. B {\bf 19} (1979), 1855--1866
\bibitem{nel2}
         D.~R.~Nelson,
         \textit{Defects in Superfluids, Superconductors, and Membranes},
         In: ``Fluctuating Geometries in Statistical Mechanics and Field Theory'', Les Houches Proceedings,
         Session LXII, 1994 (NATO ASI Series vol 62), 423--478 {\sf ArXiv:\,cond-mat/9502114}
\bibitem{kl11}
         H. Kleinert, {\it Gauge Fields in Condensed Matter. Superflow and Vortex Lines}.
         Vol. I (World Scientific, Singapore, 1989)
\bibitem{kl12}
         H. Kleinert, {\it Gauge Fields in Condensed Matter. Stresses and Defects}.
         Vol. II (World Scientific, Singapore, 1989)
\bibitem{kl2}
         H. Kleinert, {\it Vortex, Defect, and Monopole Gauge Fields for
         Nambu-Goldstone Systems and their Phase Transitions},
         {\sf ArXiv:\,cond-mat/9503030}
\bibitem{yam}
         T.~Yamamoto, T.~Izuyama, J. Phys. Soc. Japan {\bf 57} (1988),
         3742--3752
\bibitem{rab}
         S.~Panyukov, Y.~Rabin, Phys. Rev. B {\bf 59} (1999-I), 13657--13671
\bibitem{zan1}
        J. Zaanen, Z. Nussinov, S. I. Mukhin, Ann. Phys. (NY)
        {\bf 310} (2004), 181--260
\bibitem{igg}
         P.~D.~Isp\'anovity, I.~Groma, G.~Gy\"orgyi,
         Phys. Rev. B {\bf 78} (2008), 024119 [10 pages]
\bibitem{saito}
         R.~Saito, G.~Dresselhaus, M. S.~Dresselhaus,
         {\it Physical Properties of Carbon Nanotubes}
         (Imperial College Press, Imperial College, London, 1998)
\bibitem{toman}
          D.~Tom\'anek, R.~J.~Enbody (Eds.), {\it Science
          and Application of Nanotubes} (Kluwer Academic Publishers,
          etc., New York, 2002)
\bibitem{g4}
         M. Yu. Gutkin, A. G. Sheinerman,
         Phys. Solid State {\bf 49} (2007), 1672--1679
\bibitem{dkl}
         J. Dietel, H. Kleinert,
         Phys. Rev. B {\bf 79} (2009), 245415
\bibitem{graph0}
     A. Carpio, L. L. Bonilla, F. de Juan, M. A. H. Vozmediano,
      New Journal of Physics {\bf 10} (2008), 053021
\bibitem{graph}
     F. de Juan, A. Cortijo, M. A. H. Vozmediano,
      Nucl. Phys. B {\bf 828} (2010), 625--637
\bibitem{graph1}
     S. Bhowmick, U. V. Waghmare,
     Phys. Rev. B {\bf 81} (2010), 155416
\bibitem{pei}
         R. E. Peierls,
         Proc. Phys. Soc. {\bf 52} (1940), 34--37
\bibitem{nab}
         F. R. N. Nabarro,
         Proc. Phys. Soc. {\bf 59} (1947), 256--272
\bibitem{brail1}
         A. D. Brailsford,
         Phys. Rev. {\bf 142} (1966), 383--387
\bibitem{kun}
         I. A. Kunin,
         {\it Theory of Elastic Media with Microstructure}
         (Nauka, Moscow, 1975) [In Russian]
\bibitem{cem1}
         A. C. Eringen, J. Appl. Phys. {\bf 54} (1983), 4703--4710
\bibitem{g1}
         M. Yu. Gutkin, E. C. Aifantis,
         Scr. Mater. {\bf 35} (1996), 1353--1358
\bibitem{g2}
         M. Yu. Gutkin, E. C. Aifantis,
         Scr. Mater. {\bf 36} (1997), 129--135
\bibitem{val}
         M. C. Valsakumar, D. Sahoo,
         Bull. Mater. Sci. {\bf 10} (1988), 3--44
\bibitem{ed1}
         D. G. B. Edelen, Int. J. Engng Sci. {\bf 34} (1996), 81--86
\bibitem{mal1}
         C. Malyshev, Ann. Phys. (NY) {\bf 286} (2000), 249--277
\bibitem{laz2}
         M. Lazar, J. Phys. A: Math. Gen. {\bf 35} (2002), 1983--2004
\bibitem{laz3}
         M. Lazar,
         J. Phys. A: Math. Gen. {\bf 36} (2003), 1415--1438
\bibitem{haif}
         A. Seeger,
         {\it The application of second-order effects
         in elasticity to problems of crystal physics}, In:
         Second-Order Effects in Elasticity, Plasticity,
         and Fluid Dynamics. Int. Symp., Haifa, Israel,
         April 23--27, 1962 (Eds., M. Reiner, D. Abir, Pergamon Press, Oxford, 1964),
         pp. 129--144
\bibitem{mal2}
         C. Malyshev, J. Phys. A:  Math. Theor. {\bf 40} (2007),
         10657--10684
\bibitem{pop3}
          V. N. Popov, \textit{Functional Integrals in Quantum Field
           Theory and Statistical Physics} (D. Reidel, Dordrecht,1983)
\bibitem{pop4}
          V. N. Popov, \textit{Functional Integrals and Collective
          Excitations} (Cambridge University Press, Cambridge, 1987, 1990)
\bibitem{riv}
          R. J. Rivers, \textit{Path Integral Methods in Quantum Field Theory} (Cambridge University Press, Cambridge, 1987)
\bibitem{yr}
         V. N. Popov, V. S. Yarunin: \textit{Collective Effects in Quantum Statistics of Radiation and Matter} (Kluwer, Dordrecht, 1988)
\bibitem{kl3}
          H. Kleinert, \textit{Path Integrals in Quantum Mechanics,
          Statistics, Polymer Physics, and Financial Markets} (World
          Scientific, Singapore, 1990, 1995, 2004)
\bibitem{zinn}
           J. Zinn-Justin, {\it Quantum Field Theory and Critical Phenomena} (Clarendon Press, Oxford, 1996)
\bibitem{deut}
         C. Deutsch, M. Lavaud,
         Phys. Rev. A {\bf 9} (1974), 2598--2616
\bibitem{jos}
         J. V. Jos\'e, L. P. Kadanoff, S. Kirkpatrick, D. R. Nelson,         Phys. Rev. B {\bf 16} (1977), 1217--1241
\bibitem{nien}
         B. Nienhuis,
         J. Stat. Phys. {\bf 34} (1984), 731--761
\bibitem{pol}
        A. M. Polyakov, {\it Gauge Fields and Strings}
        (Harwood, London, 1987)
\bibitem{ab}
         A. A. Abrikosov (jr), Ya. I. Kogan, Zh. Eksp. Teor. Fiz.
         {\bf 96} (1989), 418--436
\bibitem{ab1}
         A. A. Abrikosov (jr), Ya. I. Kogan, Int. J. Mod. Phys. A
         {\bf 6} (1991), 1501--1524
\bibitem{teod}
         C. Teodosiu,
         {\it Elastic Models of Crystal Defects}
         (Springer--Verlag, Berlin, etc., 1982)
\bibitem{hirth}
         J. P. Hirth, J. Lothe,
         {\it Theory of Dislocations} (Wiley, New York, etc., 1982)
\bibitem{vol}
         M. O. Katanaev, I. V. Volovich, Ann. Phys. (NY) {\bf 216} (1992), 1--28
\bibitem{sard}
         G. Sardanashvily,
         Theor. Math. Phys. {\bf 132} (2002), 1163--1171
\bibitem{hehl}
         F. W. Hehl, Y. N. Obukhov,
         Ann. Fond. L. de Broglie {\bf 32}
         (2007), 157--194
\bibitem{katan}
         G. de Berredo-Peixoto, M. O. Katanaev, {\it Tube} {\it Dislocations} {\it in} {\it Gravity},
         {\sf ArXiv:} {\sf 0810.0243}
\bibitem{kl4}
         H. Kleinert, J. Zaanen, Phys. Lett. A {\bf 324} (2004), 361--365
\bibitem{dv}
         R. de Wit, {\it Linear theory of static disclinations},
         In: Fundamental Aspects of Dislocation, Nat. Bur. Stand. (US), Spec. Publ. 317, vol. 1, 1970 (Eds., J. A. Simmons, R. de Wit, R. Bullough), pp. 651--673
\bibitem{dv1}
         R. de Wit, J. Res. Natl. Bur. Stand. Sect. A {\bf 77} (1973), 49--100; 359--368; 607--658
\bibitem{lh}
         E. Kr\"oner, {\it Continuum theory of defects}, In:
         Physique des D\'efauts, Les Houches, Session XXXV, 1980 (Eds., R. Balian, {\it et al.}, North--Holland, Amsterdam, 1981), pp.
         215--316
\bibitem{pss}
         M. Lazar, G. A. Maugin, E. C. Aifantis, Phys. Stat. Sol. (b) {\bf 242} (2005), 2365--2390
\bibitem{colem}
         S. Coleman, Phys. Rev. D {\bf 11} (1975), 2088--2097
\bibitem{sam}
         S. Samuel, Phys. Rev. D {\bf 18} (1978), 1916--1932
\bibitem{frohl}
         J. Fr\"ohlich, T. Spencer, J. Stat. Phys. {\bf 24} (1981), 617--701
\bibitem{mond}
          L. Mondaini, E. C. Marino, J. Stat. Phys. {\bf 118} (2005), 767--779


\end{thebibliography}
\end{document}